\documentclass[conference]{IEEEtran}
\IEEEoverridecommandlockouts
\usepackage{cite}
\usepackage{amsmath,amssymb,amsfonts}
\usepackage{algorithmic}
\usepackage{graphicx}
\usepackage{textcomp}
\usepackage[dvipsnames]{xcolor}
\usepackage{tcolorbox}
\usepackage{url}
\usepackage[T1]{fontenc}
\usepackage[implicit=false,hidelinks]{hyperref}
\def\BibTeX{{\rm B\kern-.05em{\sc i\kern-.025em b}\kern-.08em
    T\kern-.1667em\lower.7ex\hbox{E}\kern-.125emX}}

\begin{document}

\title{Bug-Fix Variants: Visualizing \\ Unique Source Code Changes across GitHub Forks}

\author{\IEEEauthorblockN{Daigo Imamura, Takashi Ishio, Raula Gaikovina Kula, Kenichi Matsumoto}
\IEEEauthorblockA{\textit{Graduate School of Science and Technology} \\
\textit{Nara Institute of Science and Technology}\\
Ikoma, Nara, Japan \\
\{imamura.daigo.ia0, ishio, raula-k, matumoto\}@is.naist.jp}
}

\maketitle

\begin{abstract}
Forking is a common practice for developers when building upon on already existing projects. 
These forks create variants, which have
a common code base but then evolve the code in different directions, which is specific to that forked project requirements. 
An interesting side-effect of having multiple forks is the ability to select between different evolution directions of the code which is based on developers fixing bugs in the code base.
However, the key issue that this decentralized form of information is difficult to analyze.
In this study, we propose a visualization to analyze active changes in fork repositories that have not been merged back to the original project.
Our visualization shows code commit activities in multiple forks with highlight on bug fix commits in the history of forks.  
While the commit activity of each repository is visualized similarly to the code frequency view of GitHub, our view shows only commits unique to fork repositories.
To illustrate the effectiveness of our visualization, we have applied our view to two use cases: identifying forks from a repository no longer maintained, and identifying a bug fix among forks. 
In the first case, we identify a fork of a suspended project named Obfuscator-LLVM.
Our view shows the original repository and its most active fork that continue the development on the top. 
In the second case, we identify a bug fix in a fork of Clipy project.
Our view shows that the most active fork has its own bug fixes; we could easily identify a patch for the bug highlighted in the view.
As a new ideas paper, we then present our outline of three research questions to spark real world use-cases and goals for our  visualization has the potential to uncover.  
A prototype of our visualization is available at \textcolor{blue}{\url{https://naist-se.github.io/vissoft2022/}}
\end{abstract}

\section{Introduction}

Forking of free and open-source repositories has become a popular practice in software development.
There are two reasons why developers create forks.
The first is so software developers can contribute to an existing projects by creating their own copy of a repository, then later send pull requests to the original repository~\cite{gousios_exploratory_2014} (i.e., social forks).
The second reason is when software developers may create forks (i.e., hard forks) to release their own variant of the project that are independently developed of the original project~\cite{rastogi_forking_2016}.  
Some hard forks are created to revive dead projects.
There forks are created as noncompetetive alternatives to the original projects~\cite{zhou_how_2020}.  

While forking provides a convenient way for developers to introduce new features and fix bugs, the decentralization generates information that is widespread and not easy to process, analyze and summarize.
For instance, Stănciulescu et al.~\cite{stanciulescu_forked_2015} reported a case that the main repositry did not include useful features shared in multiple forks.
Hata et al.~\cite{hata_same_2021} also reported that individual repositories modify cloned source files in different ways for their own purposes.
Ren et al.~\cite{ren_identifying_2019} analyzed the duplicate development caused by unawareness of activities in other forks. 
Finding useful changes in the multitude of forks is extremely hard for the community members~\cite{stanciulescu_forked_2015}.

Prior fork visualizations exist, but it is difficult to compare source code changes against the whole set of forks.
For example, GitHub provides a network graph~\cite{dabbish_social_2012}.
However, the view does not work for a large number of forks as it tries to show all forks recorded in GitHub~\cite{zhou_identifying_2018}.
Although fork summary generation techniques~\cite{zhou_identifying_2018,zhang_forkxplorer_2021} have been proposed to extract a list of features added to a fork, the users still need to choose a small number of forks to be analyzed by the methods.
GitHub provides the function to view the commit history of a single repository, but does not support the comparison of multiple repositories.

In this study, we propose a visualization to provide an overview of development activities in fork repositories.
This is so that users can analyze i) useful forks from revived projects, ii) active hard forks including additional features and iii) bug fixes that have not been merged to the original project.
To automatically select a small number of forks, we measure the number of \emph{unique commits} in each fork.
Our key idea is to filter out noise (i.e., social forks or those not related to fixing bugs) in forked repositories \cite{zhou_identifying_2018}. 
We assume that hard forks are likely to develop their versions by fixing bugs.  Hence, they are more likely to have unique commits not included in other forks.

We have implemented our visualization as a prototype; the tool takes as input a GitHub repository URL and generates a HTML page with an interactive plot.
Users can interactively check unique commits in active forks.
As case studies, we have used the view to analyze two groups of forks; the first scenario is to identify an active fork from an archived repository.
The second scenario is to identify a bug fix in a fork that has not been merged to the original project.

In the remainder of the paper, 
Section~\ref{sec:proposal} explains our visualization. 
Section~\ref{sec:casestudy} explains the two use case scenarios.
Section~\ref{sec:conclusion} describes the current status and future work.




\begin{figure*}[t]
    \centering
    \includegraphics[width=.7\linewidth]{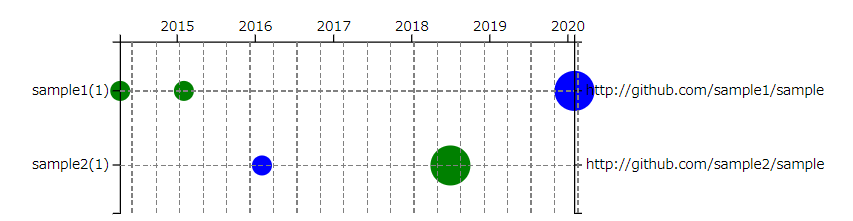}
    \caption{An example zoomed in view for two forks. The size of the bubble indicates the number of added lines by the commit in a log scale. Note that green is \textbf{\textcolor{OliveGreen}{added code}}, while blue indicates a \textbf{\textcolor{blue}{bug fix}.}}
    \label{fig:example}
\end{figure*}

\section{Our visualization}
\label{sec:proposal}

%

We propose a visualization of fork repositories to identifying forks such that revive a dead project and/or include bug fixes that have not been merged back to the original project if they exist.
Our method takes as input a project repository $R_0$ and its fork repositories $R_1, \cdots, R_n$, and provides an interactive view to investigate the contents of the fork repositories.

\subsection{Repository Selection}

It is assumed that a popular repository tends to have a huge number of forks.
However, most of them are social forks.  They include a simple copy without source code changes and a working copy whose changes have been merged to the original repository.
The remaining forks are hard forks introducing their own changes.  But many of them also become inactive eventually~\cite{zhou_how_2020}.
To distinguish active hard forks from those forks, we measure the number of divergent commits $d(R_i)$ for each fork repository defined as follows:
\begin{eqnarray*}
d(R_i) = |Commits(R_i) \setminus Commits(R_0)|
\end{eqnarray*}
where $Commits(R)$ represents a set of commits in all branches recorded in the repository $R$.
Our current implementation uses commit IDs to compare commits.
It should be noted that a pull request sent to $R_0$ is represented by commits in a branch whose name is the pull request ID in $R_0$.  
Hence, code changes in pull requests are excluded from $d(R_i)$.

If a fork includes many unique source code changes that have not been merged to the original repository, we consider the fork as important for users.
Suppose that the repositories $R_1, \cdots, R_n$ are sorted by the descending order of $d(R_i)$.
We extract unique commits from each fork as follows:
\begin{eqnarray*}
U(R_i) = Commits(R_i) ~ \setminus~ \bigcup_{j<i} Commits(R_j) 
\end{eqnarray*}
The constraint $j<i$ is to remove commits in more active forks.
We do not use $j\neq i$ because it may exclude commits that are shared in multiple forks but not in the main repository.
If $U(R_i) = \phi$, the repository is excluded from visualization. 

In the repository selection step, we prefer the number of commits to the total number of added lines of code of commits.  This is because the number of changed lines is affected by non-essential changes such as source code layout~\cite{kawrykow_non-essential_2011}.

\subsection{Visual Designs}

We selected the bubble chart as it depicts a time-series on the horizontal axis.
For the vertical axis, we placed the list of repositories.
Fig.~\ref{fig:example} shows an example of our view for two artificial repositories.
In the view, the original repository $R_0$ is always located at the top.
Its fork repositories are listed below.
As active forks are more important, the order of fork repositories is decided by $d(R_i)$.


To easily identify the repositories, we show the GitHub user name of the owner of the repository on the left side.  We use only the user name because most of forks use the same repository name as the original repository.  This is the same choice as the GitHub network graph.
On the right side, there is a full name and link to the GitHub repository page.

To enable a user to quickly check whether the repository includes a bug fix or not, we also shows the number of bug fix commits in the repository.
We detect bug fix commits using the heuristics that have been used in existing work~\cite{kim_predicting_2007,zhong_empirical_2015}.
A commit is regarded as bug fix if it includes one of keywords ``\texttt{fix}'', ``\texttt{bug}'', ``\texttt{error}'', or ``\texttt{issue}'' and a number starting with a ``\texttt{\#}'' likely representing an issue ID on an issue tracking system.

Each bubble represents a unique commit.  
The vertical position indicates that the repository includes the unique commit.
The horizontal position indicates the timestamp of the commit.
The size of the bubble is the number of added lines of code.
We ignore the number of deleted lines because we assume that commits just deleting code unlikely introduces a new feature.
To limit the size of a bubble, we use a logarithmic function.
To enable users to identify bug fix commits, we use two colors, green and blue, for bubbles.
A green bubble represents a regular commit because the GitHub code frequency view uses green to show the number of added lines.
A blue bubble represents a bug fix commit.

\subsection{Interactivity}
Unfortunately, we are unable to depict the interactive features in the paper.
For interactive purposes, we enable users to filter bug fix commits in the view.
A user can see a commit message by hovering on a bubble.
Furthermore, clicking on a bubble opens the commit page on GitHub so that a user can see actual source code changes performed by the commit.
Upon acceptance, we plan to release an online interactive demonstration for readers to explore.


\begin{figure*}[t]
    \centering
    \includegraphics[width=.8\linewidth]{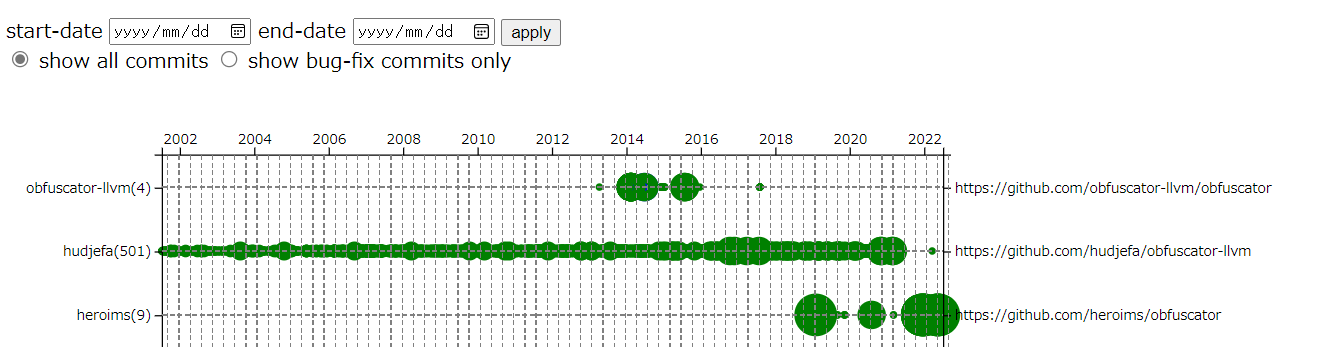}
    \caption{Case 1: Visualization of the Obfuscator-LLVM project.  This figure shows only the top 3 repositories of the view due to space limitation.}
    \label{fig:obfuscator}
\end{figure*}

\subsection{Implementation}

We have implemented our prototype comprising two components: analyzer and visualizer.
The analyzer component takes as input a GitHub repository URL specifying $R_0$. 
The tool obtains a list of fork repositories ($R_1, \cdots, R_n$) using GitHub API.  The list includes not only direct forks but also forks of forks.
Then, we download all commit IDs and messages in the repositories and computes $d(R_i)$ and $U(R_i)$ for each repository.
For each unique commit, we calculate the number of added lines to source files as the change size using the \texttt{git} command.
As the command regards the change size of a merge commit as zero, merge commits are excluded from visualization.
To extract the change size, the current version of the tool makes local clones of fork repositories including unique commits.
While a user can scroll down a large view, 
our implementation analyzes and shows only the top-10 forks by default to reduce both the size of a view and the analysis time.
The analysis result is saved in a JSON format.

The visualizer component is a graphical user interface working on \texttt{D3.js}.
It loads the JSON file generated by the analyzer and then produces an interactive bubble chart in the SVG graphic format that can be manipulated on a web browser.
The interactive view provides a filtering feature to show only bug-fix commits.
In addition, the view allows users to specify a time period to be visualized.

\section{Case Studies}
\label{sec:casestudy}
We now present two cases where our visualization can be used to identify useful forks, with potential interest to the user.

\subsection{Case 1. Identifying active fork repositories}
In the first case study, suppose that we need to search for active forks of a repository that is inactive by both maintainers or contributors.
The project obfuscator-llvm~\cite{ieeespro2015-JunodRWM} is a code obfuscation tool that works at source-code level independently of the programming language and of the target architecture.  While it is dependent on (and works for) the LLVM compilation suite~\cite{1281665}, the official repository supports only LLVM 3.3 through 4.0.
The repository has 994 forks as of June 2022.
66 out of 994 include at least one commit unique to the fork repositories.

Fig.~\ref{fig:obfuscator} shows our view of the project.
The top repository is the original repository.
As indicated, the original repository is no longer being updated.
The second repository includes a large number of unique commits.  
Unfortunately they are not development activities of the owner; 
the repository imported commits from its dependencies (i.e., LLVM 12) so that the repository owner can release a precompiled binary\footnote{\url{https://github.com/hudjefa/obfuscator-llvm/releases/tag/v12.0.0}}.
The third repository\footnote{\url{https://github.com/heroims/obfuscator}} is the active repository that updates the obfuscation tool to support LLVM 5 and later.
The timeline shows that the repository has continued the development as of 2022.

In addition to the active fork, we found several variants in the view, although they are omitted from Fig.~\ref{fig:obfuscator} due to space limitation.
The fourth repository\footnote{\url{https://github.com/dinghaowu/obfuscator}} has a unique branch named \texttt{dop-llvm}; according to the first commit message on the branch, the branch is for Dynamic Opaque Predicate Obfuscator.
Similarly, the fifth repository\footnote{\url{https://github.com/kaikj/obfuscator}} has a branch named \texttt{data-flow-transformation}.
The branch includes many small commits within a short period.
Although we could not identify the usage of the version, the commit messages say that the branch is to implement pointer manipulation and control-flow obfuscation.
The sixth repository\footnote{\url{https://github.com/spelle/obfuscator}} imported LLVM and changed build files.
The seventh repository is forked from the sixth repository and further changed the build process.  The repository releases LLVM Obfuscator Pass with its own README file.  It says that the fork does not require origin LLVM codes for compiling\footnote{\url{https://github.com/RyanKung/obfuscator/blob/llvm-11/README.md}}.
The eighth repository imported LLVM 9.0 to its \texttt{llvm\_9.0-dev} branch\footnote{\url{https://github.com/Magic-King/obfuscator/tree/llvm_9.0-dev}}.
The branch is likely for debugging because the README file on the branch describes the information of a bug including problematic command line options and a log of the case.
The nineth repository has few commits to update for Apple M1\footnote{\url{https://github.com/lemon4ex/obfuscator/commits/apple-llvm-20210628}}.
The tenth repository is a cherry-picking repository that imported only a certain pull request that has been merged to the original repository\footnote{\url{https://github.com/mbazaliy/obfuscator/commits/llvm-3.5}}. 

While analyzing 994 forks of the original repository using the raw list of forks on GitHub are unrealistic, our view enabled us to investigate only the most active forks. 
We find interesting changes in the active forks; 
the clues to understand the changes in those fork repositories are their commit messages, branch names, and README files representing the intention of the fork owner.
As our view shows unique commits on the timeline, we could easily open the corresponding web pages on GitHub.   The commit web pages including the branch names and change details helped us to investigate changes performed by the fork owners.


\begin{figure*}[t]
    \centering
    \includegraphics[width=.85\linewidth]{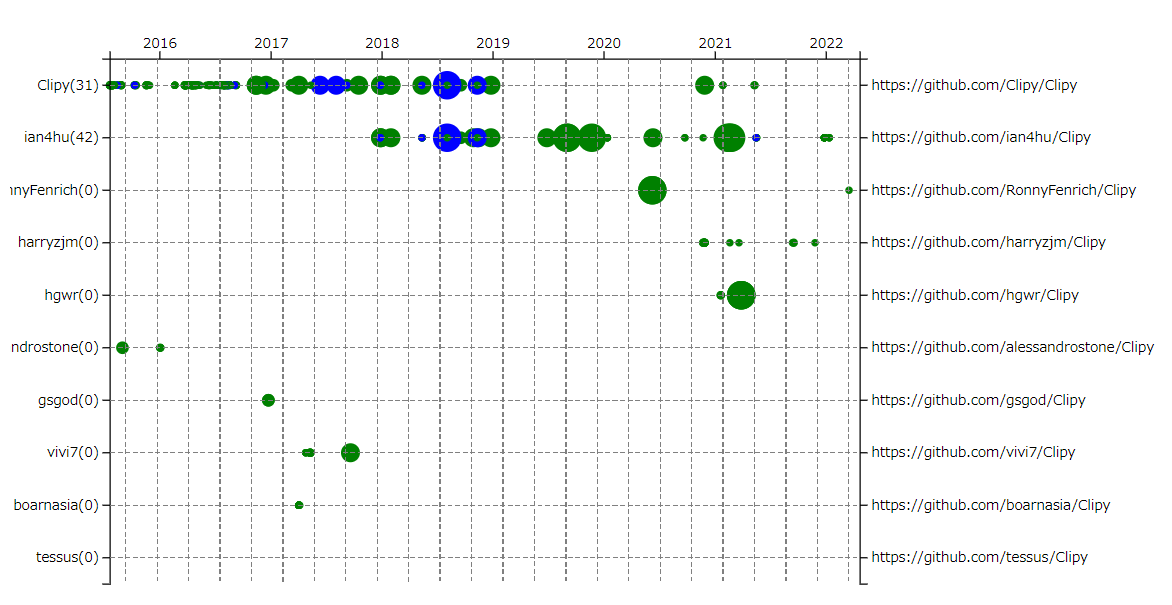}
    \caption{Case 2: Visualization for the Clipy project}
    \label{fig:clipy}
\end{figure*}

\begin{figure*}[t]
    \centering
    \includegraphics[width=.9\linewidth]{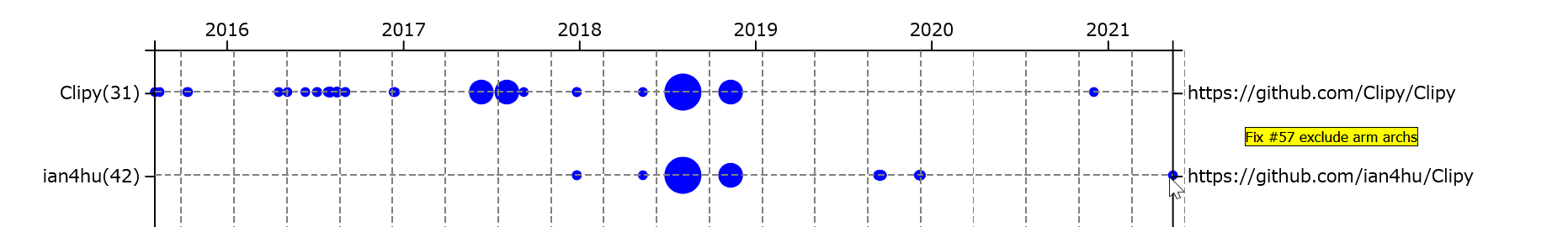}
    \caption{Our view has a filtering feature to show only bug fix commits.  Hovering on a commit shows that its commit message as tooltip.  A click on the circle opens its corresponding commit page on GitHub.}
    \label{fig:clipy-fix}
\end{figure*}
\begin{tcolorbox}
\textbf{Key Finding:}
Using our visualization we are able to identify the most active forks.
After a deeper investigation, these active forks had interesting changes related to the commit messages, branch names and README files. 
\end{tcolorbox}





\subsection{Case 2. Identifying a patch in a fork repository}

In this example, consider the case of searching a repository that has solved a particular problem.
There is a repository of tools to enhance the Mac clipboard function called Clipy\footnote{\url{https://github.com/Clipy/Clipy}}.
This program has been reported to have a problem that prevents it from starting under certain conditions, but a fork that solves this problem exists.
The latest release from the original repository is 1.2.1 on October 2018.
The original repository has updated dependent libraries but not fixed any recent issues. 
A hard fork by a user has fixed many issues and releases newer versions.  
1.2.12\footnote{\url{https://github.com/ian4hu/Clipy/releases/tag/1.2.12}} on December 2021 is available on the fork.
The issue comments in the original repository sometimes refer to the fork\footnote{\url{https://github.com/Clipy/Clipy/issues/374\#issuecomment-534025090}}.

Fig.~\ref{fig:clipy} shows our view of the project.
While the project has 487 forks, the figure shows only the top-10 out of 61 forks including at least one unique commits.
The top is the original repository.
The second repository is the active fork that has fixed many issues.
Fig.~\ref{fig:clipy-fix} shows only the bug fix commits using the filtering feature.
Some blue bubbles having the same horizontal position in the 2018-2019 period are the same bug fixes in the two repositories.
They have different commit IDs due to early unique commits in the fork.
On the other hand, some bug fixes are available only in the fork.
A bug fix commit in 2021\footnote{\url{https://github.com/ian4hu/Clipy/commit/d40ea4cf8}} is an update for a launch problem on Apple M1 processor\footnote{\url{https://github.com/ian4hu/Clipy/issues/57}}; the fix in the fork appeared in an issue discussion in the original repository\footnote{\url{https://github.com/Clipy/Clipy/issues/430}}.

The view also shows that the other forks do not have their own bug fixing commits.
For example, the third repository has some recent unique commits but they are on a branch named ``\texttt{feature/Search}.''  The commits are likely to introduce its own feature.
The fourth repository also changed code related to search function according to the commit messages.
The fifth repository has branches including a keyword \texttt{personal-build}.
The repository releases its own patched version 1.2.1.6\footnote{\url{https://github.com/hgwr/Clipy/releases/tag/build-1.2.1.6}} that is different from the second repository.
The other repositories listed in the figure are old forks as indicated in the timeline.  They are forked around five years ago and no longer maintained.

\begin{tcolorbox}
\textbf{Key Finding:}
Using our visualization we are able to identify forks that might have fixed bugs of interest.
In this case, the original repository is no longer maintained anymore.
\end{tcolorbox}




\section{Conclusion and Future Outlook}
\label{sec:conclusion}

As a new ideas paper, we propose a visualization as an alternative to summarizing and selecting important information for relevant forks.
Through two case studies, we show how our tool is able to detect any fork that is revived, and how to identify the evolution directions which a certain bug is being patched. 
On the other hand, these preliminary results raise other questions that would be interesting to get feedback.
We acknowledge that the limitation of this study is that the evaluation is incomplete.
Future work is to improve the tool so that it is scalable and useful.
Although our view can show an overview of active forks, a user still has to manually inspect them one by one.  
We would like to investigate factors to automatically identify active and useful forks, such as the number of open issues, the number of pull requests to forks, version numbers related to unique commits, or the timestamp of the last commit.
We also plan to conduct a comprehensive empirical study to strengthen our claims. 

Our goal is to foster discussion among the visualization community that leads to collaborations between other research groups to adopt, and expand our visualization idea.
Hence, we outline the following research questions that this study might spark as follows:
\begin{itemize}
    \item \textit{How can this visualization complement or compare to using textual summarizing techniques?} In future work, integrating a summary generation technique for forks \cite{zhou_identifying_2018,zhang_forkxplorer_2021} would be effective to improve the usefulness of the visualization.

    \item \textit{How can we use clone-and-own techniques to expand from not only bug-fixing, but other kinds of techniques to evolve the code?} We are also interested in clone-and-own detection techniques such as \cite{ishio_source_2017,kawamitsu_identifying_2014} to analyze forks that are unrecorded in GitHub.

    \item \textit{Can we leverage maintainer or contributor information (social data), to present some interesting evolution from a socio-technical perspective?} 
    Similar to tracking the life and death of software ecosystems   \cite{kula2019}, another direction is taking human factors into account, e.g. analyzing how forks are supported by the community.
\end{itemize}


\section*{Acknowledgment}

This work is supported by JSPS KAKENHI Grant Numbers JP20H05706 and JP18H04094.

\bibliographystyle{IEEEtranS}
\bibliography{reference}

\end{document}